\documentclass[aip,jap,preprint,superscriptaddress,showpacs]{revtex4-1}
\usepackage{amssymb}
\usepackage{geometry}                 
\usepackage{graphicx}
\usepackage[utf8]{inputenc}
\usepackage[english]{babel}

\newcommand{\mm}[1]{\overline{\overline{\mu}}_{#1}}
\newcommand{\mq}[1]{\overline{\overline{#1}}}

\begin{document}

\title{Frequency-dependent effective permeability tensor of unsaturated polycrystalline ferrites}
\author{Pascal Thibaudeau}
\email{pascal.thibaudeau@cea.fr}
\affiliation{CEA DAM/Le Ripault, BP 16, F-37260, Monts, FRANCE}

\author{Julien Tranchida}
\email{julien.tranchida@cea.fr}
\affiliation{CEA DAM/Le Ripault, BP 16, F-37260, Monts, FRANCE}
\affiliation{CNRS-Laboratoire de Mathématiques et Physique Théorique (UMR 7350), Fédération de Recherche "Denis Poisson" (FR2964), Département de Physique, Université de Tours, Parc de Grandmont, F-37200, Tours, FRANCE}

\date{\today}                                           

\begin{abstract}
Frequency-dependent permeability tensor for unsaturated polycrystalline ferrites is derived through an effective medium approximation that combines both domain-wall motion and rotation of domains in a single consistent scattering framework. Thus derived permeability tensor is averaged on a distribution function of the free energy that encodes paramagnetic states for anhysteretic loops. The initial permeability is computed and frequency spectra are given by varying macroscopic remanent field.  
\end{abstract}


\maketitle
\section{Introduction}
The determination of the frequency-dependent permeability tensor of a magnetic unsaturated gyrotropic medium such as ferrites materials is a long standing problem \cite{Rado:1953zp,Rado:1956ff,Grimes:1957tw}. It has been demonstrated useful to describe precisely this tensor in frequency to obtain novel radiation and scattering characteristics of microstrip antennas \cite{Pozar:1992gf}. To the physical point of view in dense polycrystalline materials, the frequency behavior of the permeability in the range from 100kHz to 20GHz is understood as a superposition of magnetization changes due to domain-walls motion and coherent rotations of magnetic domains \cite{Rado:1950fh,Rado:1953zp}. However, Grimes \cite{Grimes:1991lq,Grimes:1991ek} has suggested that the observed spectra may be also interpreted as multiple scattering of electromagnetic waves in these polycrystalline materials by both random magnetic and dielectric homogeneous spheres. Unfortunately this deduced interpretation has not been tested when magnetic materials, subject to external dc uniform magnetic field present an hysteresis behavior on the magnetization and changes the derived permeability. A convergence of these two alternate mechanisms was first recognized by Schlömann \cite{Schlomann:1970rw} who's derived a relationship between the isotropic permeability in a completely demagnetized state by computing the scattering of magnetic fields in the static limit by out-of-phase and concentrical cylinders of gyromagnetic single-domains. The extension of this idea to unsaturated magnetic states was conducted by Bouchaud and Zérah \cite{Bouchaud:1989hc} but the bridge between the local remanent magnetization, included as fractions of magnetic volumes of scatterers and the observed average distribution of magnetization directions of domains in the tridimensional space reminded elusive. To get into account the anisotropic nature of this scattering problem, Stroud \cite{Stroud:1975rm} was pioneer to generalize an old effective-medium approximation (EMA) for the conductivity tensor of a randomly inhomogeneous medium to treat materials consisting of crystallites of arbitrary shape and conductivity tensors of arbitrary symmetry. Derivation of non diagonal permeability tensors for magnetized granular composites has been then reported \cite{Abe:1996rm} and several models for the calculation of complex permeability of magnetic composite materials have been proposed \cite{Igarashi:1977gf,Neo:2010hb}. By using the EMA, even the influence of porosity induced by non-magnetic inclusions in magnetized heterogeneous materials has been computed \cite{Queffelec:2005th}. However, even in the quasi-static limit of frequency, the problem of taking into account dynamically the magnetic multiple-scattering of both domain-walls and single domains simultaneously in a consistent, anisotropic formulation is not achieved yet. This is the main purpose of this paper.

\section{Effective Medium Approximation for unsaturated ferrites}
Because electrical non-conducting media are considered, the ferrites are treated as uniform from the dielectric point of view. The amplitude of an ac magnetic field applied is sufficiently low to let first the transverse permeability tensor being independent of it and to generate a small local transverse oscillating magnetic field only. 

Saturated ferrites exhibit anisotropic permeability described by the well known Polder tensor \cite{Polder:1949yq}, resulting in the nonreciprocal behavior of microwaves into magnetized ferrite used for the design of circulators and isolators \cite{Helszajn:2008dq}. As any reference axis may be chosen to project the third component of an arbitrary magnetization vector, the $z$-axis is then taken for that. So in a cartesian frame and for any arbitrary direction, the magnetization vector is represented by only $2$ angles, a polar one $\theta$ and an azimutal one $\phi$. Then in this arbitrary frame, the gyromagnetic permeability tensor can be expressed in a general form \cite{Tyras:1959ve,Baden-Fuller:1987eu}. Such saturated ferrite is thus characterized by only three circular permeabilities which come from the response of the magnetization to a rotating ac magnetic field. This corresponds to diagonalize the general form of the gyromagnetic permeability tensor as first recognized by Tyras a long time ago \cite{Tyras:1959ve}. It is always possible to select a frame that puts the saturation magnetization vector along the $z$-axis and in that case, one of the third component of the permeability is equal to $1$ and the ferrite is characterized by only 2 remaining permeabilities.

For an unsaturated polycristalline ferrite, the spectral value of the 3 eigenvalues as a function of a given reduced remanent magnetization state $m=M_z/M_s\equiv\cos(\theta)$ pointing along the $z$-axis, is not so easily found. One may get this effective tensor as a statistical average of the Polder tensor in a non-interacting magnetic grains picture. This strategy was investigated \cite{Van-Vleck:1951ud}. Because the distribution of the magnetization direction (the magnetic texture) has to be given, and can change as a function of $m$, it results that the non-interacting assumption is strictly satisfied for a system consisted in an assembly of single-crystal spheres situated in a non-magnetic medium far from one another, a condition rarely observed in dense soft ferrites \cite{Schlomann:1958wo,Schlomann:1959ec}. No interaction between domains and domain-walls can be considered in such picture and this also neglects the influence of the shape of the magnetic domains and pores. So a two steps mechanism is introduced. The magnetostatic interaction between domains has to be treated first and then the resulting composite medium has to be statistically averaged.

For any value of $m$, consider that the effective $3\times 3$ permeability tensor $\mm{e}$ can be diagonalized and expressed in a circular frame by the following diagonal tensor (see Appendix \ref{appendix1} for the notations):
\begin{equation}
\label{mme}
\mm{e}=\left(\begin{array}{ccc}
\mu_e+\kappa_e&0&0\\
0&\mu_e-\kappa_e&0\\
0&0&{\mu_e}_z
\end{array}\right),
\end{equation}
where $\mu_e$ and $\kappa_e$ are the diagonal and off-diagonal components of the permeability expressed in a cartesian frame \cite{Tyras:1959ve}. ${\mu_e}_z$ is the permeability value which connects both the magnetization and the ac field along the $z$-axis.

\subsection{Two-phase model}
By using the same argument to the scale of a magnetic "grain", let us consider "up" and "down" borderless single domain grains which differ only by the direction of their magnetization vector and thus by a change of a sign. Their respective permeability tensors are
\begin{equation}
\label{mediummatrices} 
\begin{array}{ccc}
\mm{1}=\left(\begin{array}{ccc}
\mu+\kappa&0&0\\
0&\mu-\kappa&0\\
0&0&\mu_z
\end{array}\right)&\mathrm{and}&
\begin{array}{ccc}
\mm{2}=\left(
\begin{array}{ccc}
\mu-\kappa&0&0\\
0&\mu+\kappa&0\\
0&0&\mu_z
\end{array}\right)
\end{array}
\end{array}.
\end{equation} 
Now, one consider that each borderless domain can be represented with Polder expressions in frequency. Such analytical expressions of $\mu, \kappa$ and $\mu_z$ are determined from the solution of the Landau-Lifchitz equation in the small signal approximation such as :
\begin{eqnarray}
\mu&=&1+\frac{\eta-\imath\alpha\Omega}{(\eta-\imath\alpha\Omega)^2-\Omega^2},\\
\kappa&=&\frac{\Omega}{(\eta-\imath\alpha\Omega)^2-\Omega^2},\\
\mu_z&=&1,
\end{eqnarray}
with $\eta=H_k/M_s$ the reduced anisotropy field assuming an uniaxial symmetry, $\Omega=\omega/\gamma\mu_0M_s$ the reduced pulsation with $\gamma$ the gyromagnetic ratio, $M_s$ the saturation magnetization, $\mu_0$ the permeability of free space and $\alpha$ is the damping constant.

If the domain structure ("up" and "down") is distributed such as one cannot distinguish any direction in the perpendicular plane to the magnetization on a macroscopic scale, then the permeability tensor must be diagonal with respect to such a rotating field \cite{Schlomann:1970rw}. To get that, one may consider domains in the shape of infinite circular cylinders \cite{Schlomann:1970rw}. The demagnetizing tensor for a single cylinder embedded in a effective anisotropic medium in the quasi-static approximation is then
\begin{equation}
\mq{\Gamma}=\left(
\begin{array}{ccc}
\displaystyle{-\frac{1}{2\mu_e}}&0&0\\
0&\displaystyle{-\frac{1}{2\mu_e}}&0\\
0&0&0
\end{array}
\right).
\label{surface_matrix}
\end{equation}
Even when the medium is characterized by an anisotropic tensor, Eq.(\ref{gammatensor}) yields a diagonal $\mq{\Gamma}$ tensor with $\Gamma_{11}=\Gamma_{22}$ for both cylindrical and spherical crystallites. In our situation, $\mq{\Gamma}$ is invariant through any cylindrical rotation around the $z$-axis and preserves this structure in the rotating frame.

Now some EMA procedure has to be applied in order to get the effective quantities ($\mu_e, \kappa_e, {\mu_e}_z$) as functions of ($\mu, \kappa, \mu_z$) and $m$. It is well known that the EMA depends on the choice of the reference permeability \cite{Torquato:2002cs}. Among these choices, the strong-coupling interaction between domains is assumed and the symmetric self-consistent situation is so retained. In our situation, this corresponds to let equation (\ref{EMT1b}) to be zero. 
For each two-phase grain, the local magnetization of the composite medium is $m=(v_{1}-v_{2})/(v_{1}+v_{2})$ where $v_{1}$ (resp. $v_{2}$) is the volume occupied by the portion of the grain where the magnetization is pointing in the $z$-direction (resp. opposite to). Because the total volume is $v_{1}+v_{2}=V$, one can evaluate $v_{1}=V(1+m)/2$ and $v_{2}=V(1-m)/2$. Thus Eq.(\ref{EMT1b}) reads 
\begin{equation}
(1+m)[\mq{1}-\mq{\Gamma}(\mm{1}-\mm{e})]^{-1}(\mm{1}-\mm{e})+(1-m)[\mq{1}-\mq{\Gamma}(\mm{2}-\mm{e})]^{-1}(\mm{2}-\mm{e})=\mq{0}.\label{emt2body}
\end{equation}
By substituting matrices (\ref{mme}), (\ref{mediummatrices}) and (\ref{surface_matrix}) into equation (\ref{emt2body}), $\mu_e$, $\kappa_e$, ${\mu_e}_z$ are connected with $\mu$, $\kappa$, $\mu_z$ and $m$ as
\begin{eqnarray}
\mu_e^2&=&\mu^2\frac{\mu^2-\kappa^2}{\mu^2-m^2\kappa^2}\label{mubar}\\ 
\kappa_e&=&m\kappa\frac{\mu_e}{\mu}\label{kappabar}\\
{\mu_e}_z&=&\mu_z\label{muz2}
\end{eqnarray}

Remarkably when $m=\pm 1$, $\mu_e=\mu$, $\kappa_e=\kappa$ and ${\mu_e}_z=1$, which implies that the Polder tensor is recovered to the saturation limit. Equation (\ref{mubar}) and (\ref{kappabar}) agree with equations (3a) and (3b) of reference \cite{Bouchaud:1989hc} which state that for any cylindrically symmetric but otherwise arbitrary configuration, this EMA procedure does not depend upon the details of the domain configuration \cite{Keller:1964tw,Schlomann:1970rw}. Equation (\ref{kappabar}) also agrees with several works \cite{Rado:1953zp,Rado:1956ff,Grimes:1957tw} which relies linearly the static magnetization along the $z$-axis with the value of the off-diagonal permeability, at least for small values of $m$. 

As a consequence, it is worth noting that for an effective two-components medium and any diagonal surface tensor $\mq{\Gamma}$, the demagnetized situation $m=0$ always gives $\kappa_e=0$ and the following implicit equation for ${\mu_e}$ :
\begin{equation}
\kappa_e{\mu_e}^2\Gamma_{22}({\mu_e},{\mu_e}_z)+(1-2\mu\Gamma_{22}({\mu_e},{\mu_e}_z)){\mu_e}+\Gamma_{22}({\mu_e},{\mu_e}_z)(\mu^2-\kappa^2)-\mu=0, \label{gene2body}
\end{equation} 
depends only on the single component $\Gamma_{22}=\Gamma_{11}$. This result is a generalization of Eq.(\ref{mubar}) to any geometry and anisotropic embedded medium resulting in a diagonal surface tensor. By inserting Eq.(\ref{surface_matrix}) into Eq.(\ref{gene2body}), the equation (\ref{mubar}) is recovered.


\subsection{Three-phase model}
Neglecting any magnetic after-effect or magnetic viscosity \cite{Chikazumi:1997xy}, the permeability spectra are described by two types of magnetizing processes: gyration of domains and domain-wall motion \cite{Smit:1959df} which are generally analyzed separately \cite{Naito:1970kc}. In order to take into account the lower frequency response of the previously introduced biphasic medium, the permeability response for a domain-wall has to be consistently added. Its spectral response is commonly assumed to be modeled by an isotropic expression of permeability \cite{Tsutaoka:1995pt,Tsutaoka:1997la,Tsutaoka:2003hl,Su:2008if,Fiorillo:2009it} that follows from a stiff and dampened 180$^\mathrm{o}$ domain-wall motion \cite{Smit:1959df} :
\begin{equation}
\mu_{dw}=1+\frac{\chi_{dw}\eta_{dw}^{2}}{\eta_{dw}^2-\Omega^2-\imath\alpha_{dw}\eta_{dw}\Omega}.
\end{equation}
Here $\eta_{dw}=H_{dw}/M_s$ is the reduced domain-wall resonance field, $\alpha_{dw}$ is a damping factor and $\chi_{dw}$ represents its static susceptibility. To allow the spectral representation of such permeability to be reduced to two parameters only, its frequency permeability may be approximated as a single uniaxial magnetic domain distribution in the case of small damping and if $\chi_{dw}\approx 1/\eta_{dw}$. In order to let the static susceptibility of the domain-wall to be higher than the value of a magnetic domain, it is assumed that $\eta_{dw}$ returns a smaller value of the anisotropy field in the domain-wall than the anisotropy field $\eta$ characterizing a domain.  

\begin{figure}[thp]
\resizebox{0.9\columnwidth}{!}{\includegraphics{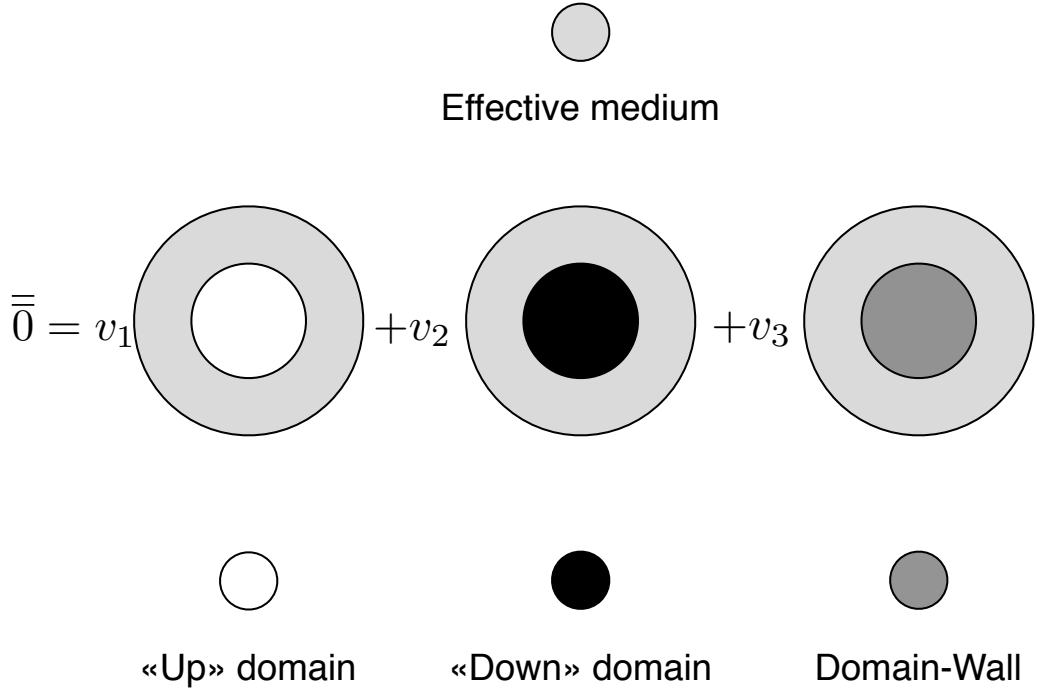}}
\caption{Schematization of the homogenization procedure for a composite medium made up of uniaxial anisotropy domains and domain-wall. \label{Fig_Scheme_EMT}}
\end{figure}

The goal is to treat on the same footing the dynamics of interacting domains and domain-wall motion for a given unsaturated value of magnetization $m$. An EMA is thus constructed with a three components system. "Up" and "down" domains are considered supplemented by a domain-wall, each component embedded in an anisotropic effective medium with a cylindrical geometry. The homogenization procedure is depicted for a composite medium made up of uniaxial anisotropy particles on Fig. (\ref{Fig_Scheme_EMT}).
This gives the following implicit equation for $\mm{e}$
\begin{equation}
\sum_{i=1}^{3}v_i\left[\mq{1}-\mq{\Gamma}_i(\mm{i}-\mm{e})\right]^{-1}(\mm{i}-\mm{e})=\mq{0},\label{EMT3}
\end{equation}
where $\mm{1}$ and $\mm{2}$ are the permeability tensors of domains and $\mm{3}\equiv\mu_{dw}\mq{1}$. Here $v_i$ are the volume fraction of each component. As a consequence of the cylindrical geometry, the same surface matrix Eq.(\ref{surface_matrix}) is taken into account for all the embedded media. The fraction volumes have to be normalized according to $\lim_{m\rightarrow\pm 1}v_3=0$ and one assumes
\begin{equation}
v_1\equiv\frac{1+m}{2(1+a(1-m^2))}, v_2\equiv\frac{1-m}{2(1+a(1-m^2))}, v_3\equiv\frac{a(1-m^2)}{1+a(1-m^2)},
\end{equation}
where $a$ describes some proportion of domain-walls in the whole system. When $a=0$ the previous two-phase situation is recovered and the limit case $a\rightarrow\infty$ provides $\mu_e=\mu_{dw}$. Because $\Gamma_{33}$ is here zero, the $zz$-component of equation (\ref{EMT3}) gives exactly ${\mu_e}_z$ and reads
\begin{equation}
{\mu_e}_z=\frac{\mu_z+a\mu_{dw}(1-m^2)}{1+a(1-m^2)}.\label{new3bodyz}
\end{equation}
One shows that ${{\mu_e}}_z=\mu_z$ when $a=0$ or $m=\pm1$ as expected. In the demagnetized state $m=0$, ${{\mu_e}}_z$ is the volume addition of both the permeability of the domain and domain-wall by reason of the superposition of the $z$-component of the magnetic induction in all media. When $\kappa_e=0$, ${\mu_e}$ is the solution of a single equation in (\ref{EMT3}) which is written directly in a third-order polynomial form
\begin{equation}
{\mu_e}^3+\frac{\mu_{dw}(1-a)+2a\mu}{1+a}{\mu_e}^2-\frac{(\mu^2-\kappa^2)(1-a)+2a\mu_{dw}\mu}{1+a}{\mu_e}-\mu_{dw}(\mu^2-\kappa^2)=0.\label{mu3}
\end{equation}
The 3-body composite medium appears isotropic but the value of ${\mu_e}$ depends surprisingly on the off-diagonal $\kappa$ term. Among all the three complex-valued solutions of equation (\ref{mu3}), the unique root with the positive imaginary part is followed continuously in frequency to get the physical picture of a lossy material. 

For $m\neq 0$, the $xx$ and $yy$-components of equation (\ref{EMT3}) expand as
\begin{equation}
\left\{
\begin{array}{rcl}
\displaystyle{{(1+m)\frac {\mu-\kappa-{\mu_e}+{\kappa_e}}{{\mu_e}+\mu-\kappa+{\kappa_e}}}+(1-m){\frac {\mu+\kappa-{\mu_e}+{\kappa_e}}{{\mu_e}+\mu+\kappa+{\kappa_e}}}+(1-m^2){\frac {2a \left( \mu_{dw}-{\mu_e}+{\kappa_e} \right) }{{\mu_e}+\mu_{dw}+{\kappa_e}}}}&=&0\\
\displaystyle{{(1+m)\frac {\mu+\kappa-{\mu_e}-{\kappa_e}}{{\mu_e}+\mu+\kappa-{\kappa_e}}}+(1-m){\frac {\mu-\kappa-{\mu_e}-{\kappa_e}}{{\mu_e}+\mu-\kappa-{\kappa_e}}}+(1-m^2){\frac {2a \left(\mu_{dw}-{\mu_e}-{\kappa_e} \right) }{{\mu_e}+\mu_{dw}-{\kappa_e}}}}&=&0
\end{array}
\right.\nonumber
\label{new3body}
\end{equation}
and have to be solved simultaneously to get $\mu_e$ and $\kappa_e$. These equations are identical by interchanging $\kappa\leftrightarrow -\kappa$ and $\kappa_e\leftrightarrow -\kappa_e$ which guaranties an exhibition of the quantities $\kappa^2$, $\kappa_e^2$ and $\kappa\kappa_e$. A bi-dimensional root-finding numerical algorithm in the complex plane using a variant of the Newton procedure is developed to get the couple $({\mu_e}, \kappa_e)$ for any values of $m$, $a$, $\mu$, $\kappa$ and $\mu_{dw}$. As for the two-phase situation, Eqs.(\ref{new3bodyz}) and (\ref{new3body}) show that if $m=\pm 1$ then $\mu_e=\mu$, $\kappa_e=\kappa$ and ${\mu_e}_z=1$. This is also true by construction of the three-phase model. This implies that $\mm{e}$ goes to the spectral Polder tensor to the saturation limit.

\subsection{Anhysteretic texture}
The presence of a large amount of such composite media not necessarily distributed at random, has to be assessed by a space dependent fluctuation in the magnetization direction to represent its magnetic texture. The statistical average of any function $U(\theta,\phi)$ of the direction angles $\theta$ and $\phi$ is defined as the following weighted normalized integral \cite{Van-Vleck:1951ud}:
\begin{equation}
\langle U\rangle\equiv\int_0^{2\pi}d\phi\int_0^\pi d\theta\sin\theta f(\theta,\phi)U(\theta,\phi),
\end{equation}
where $f(\theta,\phi)$ is a normalized texture function that distribute the angles on the unit-sphere at equilibrium and is a function of external parameters such as applied dc magnetic field, constrains etc.

When a full demagnetized state is desired at the macroscopic level, then the average remanent field $\langle m\rangle$ should remain zero but on a statistical distribution of the local direction of the magnetization. It is particularly true for the uniform angular distribution. It is noted that even if $\kappa_e\neq 0$, one could also realized $\langle\kappa_e\rangle=0$ for any angular distribution $f(\theta,\phi)$ of the composite medium by satisfying
\begin{equation}
\langle\kappa_e\rangle=\int_0^{2\pi}d\phi\int_0^\pi d\theta\sin\theta f(\theta,\phi)\left(\frac{{\mu_e}_z-{\mu_e}}{2}\sin^2\theta\sin 2\phi+\kappa_e\cos\theta\right)=0,
\end{equation} 
which is true when $f$ is uniform. This means that if desired, a locally anisotropic composite medium which behaves statically isotropic may be constructed, once their composite media are properly distributed. 

For an uniform distribution of composite domains (2- or 3-boby) in the demagnetized state, one has $\langle\kappa_e\rangle=0$ and
\begin{eqnarray}
\langle{\mu_e}\rangle&=&\int_0^{2\pi}d\phi\int_0^\pi d\theta\sin\theta \frac{1}{4\pi}\left({\mu_e}+({\mu_e}_z-{\mu_e})\sin^2\theta\cos^2\phi\right)=\frac{2{\mu_e}+{\mu_e}_z}{3},\\
\langle{\mu_e}_z\rangle&=&\int_0^{2\pi}d\phi\int_0^\pi d\theta\sin\theta \frac{1}{4\pi}({\mu_e}_z-({\mu_e}_z-{\mu_e})\sin^2\theta)=\frac{2{\mu_e}+{\mu_e}_z}{3},
\end{eqnarray}
for biphasic cylindrical inclusions uniformly distributed into space and embedded in a gyrotropic effective medium. So the statistical average effective tensor of a demagnetized ferrite is  
\begin{equation}
\langle\mm{e}\rangle(m=0)=\frac{1}{3}{\rm Tr}(\mm{e})\overline{\overline{1}},
\end{equation}
which is the result found by Schlömann \cite{Schlomann:1970rw}.

For an unsaturated medium when $m$ increases in the $z$-direction, the distribution of the local magnetization angles cannot be uniform and is distributed around the $z$-axis. Moreover whatever the value of $m$ may take, $\langle\kappa_e\rangle=0$ for an uniform distribution, which is in contradiction to experiments.

This problem can be circumvent by recalling that the probability that the local magnetization occupies an infinitesimal solid angle in a given direction is therefore strongly dependent on the magnetic free energy in this direction. Let
\begin{equation}
\langle m\rangle=\int_0^{2\pi}d\phi\int_0^\pi d\theta\sin\theta f(\theta,\phi)\cos\theta,
\end{equation}
the statistical average of the remanent magnetization. So when all the collective media are in the paramagnetic state driven by the exterior constant induction field $\mu_0H$ pointing in the $z$-direction, the function $f$ is then given by an equilibrium Boltzmann distribution $f(\theta,\phi)=X\exp(X\cos(\theta))/4\pi\sinh(X)$ where $X=A_sM_s\mu_0 H$ which naturally breaks the space isotropy when $H\neq 0$ \cite{Armstrong:1997gf}. The local magnetic moments of the composite medium do not follow nor instantly nor spatially, the direction of the external induction by coherent rotations but instead in defocusing their local magnetization about the average direction. The free energy is scaled by the Armstrong's parameter $A_s$ which narrows statistically the distribution $f$ when both the nature and degree of disorder are known. An estimation of this parameter is given by equating linearly $\langle m\rangle$ to $H$ for small value of $X$ and one has in our case
\begin{equation}
A_s\approx3\frac{\langle {\mu_e}_z\rangle-1}{\mu_0M_s^2},
\end{equation} 
a result also found previously \cite{Daniel:2008eu}.

Now the statistical average of $\mm{e}$ has to be derived. By regrouping all the terms, one can link exactly $\langle{\mu_e}\rangle$, $\langle\kappa_e\rangle$ and $\langle{\mu_e}_z\rangle$ with $\langle m\rangle$ to read
\begin{eqnarray}
\langle{\mu_e}\rangle&=&{\mu_e}+\frac{\langle m \rangle}{{\cal{L}}^{-1}(\langle m\rangle)}({\mu_e}_z-{\mu_e})\label{emubar}\\
\langle\kappa_e\rangle&=&\langle m \rangle\kappa_e\\
\langle{\mu_e}_z\rangle&=&{\mu_e}_z-\frac{2\langle m \rangle}{{\cal{L}}^{-1}(\langle m\rangle)}({\mu_e}_z-{\mu_e})\label{emuzbar}
\end{eqnarray} 
where ${\cal{L}}(x)=\coth(x)-1/x$ is the Langevin function and ${\cal{L}}^{-1}(x)$ is its inverse function such as ${\cal{L}}^{-1}\circ{\cal{L}}(x)={\cal{L}}\circ{\cal{L}}^{-1}(x)=x$. These three equations constitute the central result of this paper. One recognizes that $\langle m\rangle={\cal{L}}(A_s\mu_0M_sH)$ is the well known paramagnetic result. Moreover the external field is modulated by the Armstrong parameter and its contribution to rotate the average magnetization depends on the susceptibility of the material. In a general situation, both ${\mu_e}$ and $ \kappa_e$ have to be evaluated by considering a direct map between the local remanent state $m$ of the composite medium and the macroscopic remanent magnetization $\langle m\rangle$ in a some sort of mean-field approximation, i.e. $m=\langle m\rangle$. For practical applications, an approximation of the inverse Langevin function is taken \cite{Cohen:1991vn} and one has $\langle m\rangle/{\cal{L}}^{-1}(\langle m\rangle)\approx(1-\langle m\rangle^2)/(3-\langle m\rangle^2)$ for all values of $\langle m\rangle$. One observes that ${\rm{Tr}}(\langle\mm{e}\rangle)={\rm{Tr}}(\mm{e})$ is an invariant quantity whatever the value of $\langle m \rangle$ may take. Moreover when $\langle m\rangle=0$ then $\langle m\rangle /{\cal{L}}^{-1}(\langle m\rangle)=1/3$, $\langle{\mu_e}\rangle=(2{\mu_e}+{\mu_e}_z)/3$, $\langle{\mu_e}_z\rangle=(2{\mu_e}+{\mu_e}_z)/3$ and $\langle\kappa_e\rangle=0$ as expected. This is provided by the fact that $\lim_{H\rightarrow 0}f(\theta,\phi)=1/4\pi$.

In the previous description of the effective permeability tensor, the magnetization ratio $\langle m\rangle$ is a fixed parameter that states the magnetization along the $z$-axis. However the effective anisotropy field, which is characterized by the value of $\eta$ is assumed to be independent of such a state. When the material is probed by increasing an external dc field $H$, a continuous variation of $\langle m\rangle$ occurs. Neglecting any supplementary demagnetizing fields coming from the geometry enforced on the domains and domain-walls, the effective gyromagnetic resonance frequency of the magnetic domain shifts linearly with the amplitude of $H$ during this process. This experimental phenomenon has to be reproduced and to take it into account, the local effective field inside each uniform media has to be carried out from a magnetization law. Several strategies have been derived to get an anhysteretic or hysteric magnetization law, from coherent rotation mechanism \cite{Stoner:1991tw,Queffelec:2005th} to more elaborated models \cite{Jiles:1983kh,Bertotti:1998ay}. As a first step, for each branches along the hysteresis loop, a local equilibrium function $f(\theta,\phi)$ has to be evaluated to derive the hysteretic unsaturated spectral permeabilities. Even in the anhysteretic situation, one has at least to replace in the calculation of the local permeability tensor of domains and domain-wall, the anisotropy field by an effective field that rotates statically the magnetic moments from the easy axis of the corresponding medium. This is done in our case by adding $H/M_s$ to the anisotropy constants, hence expressed as a function of $\langle m\rangle$ only one has
\begin{eqnarray}
\eta&\rightarrow&\eta+\frac{{\cal{L}}^{-1}(\langle m\rangle)}{3(\langle{\mu_e}_z\rangle-1)},\\
\eta_{dw}&\rightarrow&\eta_{dw}+\frac{{\cal{L}}^{-1}(\langle m\rangle)}{3(\langle{\mu_e}_z\rangle-1)}.
\end{eqnarray}   
By reason of the dependance of $\langle{\mu_e}_z\rangle$ in the anisotropy fields, equation (\ref{emuzbar}) has to be solved self-consistently. This procedure is initiated first by considering known the case $\langle m\rangle=0$ as a guess of $\langle{\mu_e}_z(\Omega=0)\rangle$ and secondly, that an infinitesimal increment of $\langle m\rangle$ does not affect the previous value of $\langle{\mu_e}_z(\Omega=0)\rangle$ hence found. This procedure generates the desired permeability as a function of $\langle m\rangle$ by consecutive steps such as $\langle{\mu_e}_z(\Omega=0,\langle m\rangle+\delta\langle m\rangle)\rangle\approx\langle{\mu_e}_z(\Omega=0,\langle m\rangle)\rangle$.

\section{Results}
\begin{figure}[htb]
\resizebox{0.9\columnwidth}{!}{\includegraphics{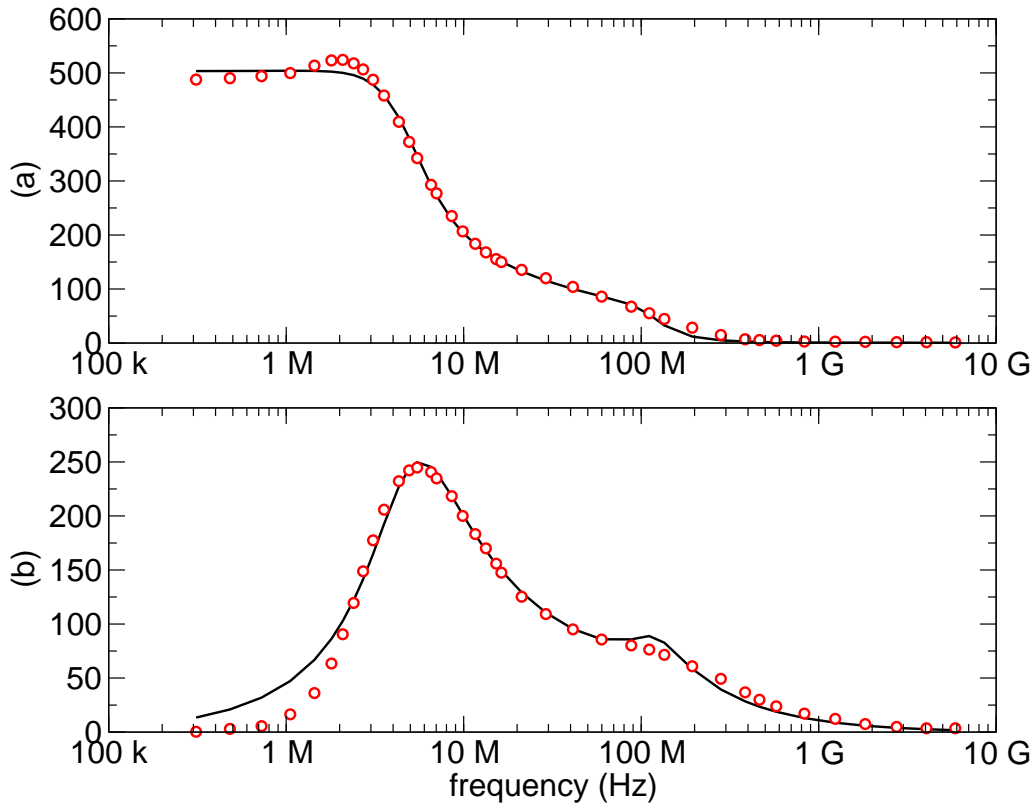}}
\caption{(Color online) Experimental real part (a) and imaginary part (b) of the effective permeability $\langle\mu_e\rangle$ measured in a coaxial wave guide at the APC7 standard for a demagnetized sintered NiZn ferrite (points). The black curve comes from the calculated effective permeability described in the text using parameters found into table \ref{param_mod}. \label{fit0}}
\end{figure}

Soft NiZn ferrites have been synthesized and the effective permeability component in a coaxial wave guide at the APC7 standard have been acquired \cite{Grimal:2006ys}. These ferrites have been demagnetized by proper thermal treatment above the Curie temperature. It has been verified by measuring hysteresis loops along and perpendicular direction to the hollow of the cylinder that the average remanent field $\langle m\rangle$ does not exceed $10^{-3}$. This spectral measurement reported on figure (\ref{fit0}) allows to fit model parameters from equation (\ref{emubar}), which are collected on table \ref{param_mod}.
\begin{table}[htb]
\begin{tabular}{ccccc}
\hline
a&$\eta$&$\alpha$&$\eta_{dw}$&$\alpha_{dw}$\\
\hline
5.0E-1&6.6E-3&2.4E-1&4.0E-4&9.6E-1\\
\hline
\end{tabular}
\caption{\label{param_mod}Parameters of the described model for demagnetized NiZn ferrites}
\end{table}
The fitted values $(\eta,\alpha)$ for domains are in agreement with reference \cite{Bouchaud:1989hc}. The ratio between domain-wall and domain frequencies is ranged from $(2-5)10^{-2}$ in MgFe ferrites \cite{Smit:1959df} which is compatible with our fitted values. It has been reported a strong dependance on composition \cite{Grimal:2006ys}, porosity \cite{Barba:2004zh}, grain size \cite{Rado:1950fh,Gelin:2005uo} and applied stresses \cite{LeFoch:1987qp} in the low frequency regime, where domain-wall motion is the supposed dominant mechanism. However, the reported dependancies may be interpreted more as a consequence of the EMA that mimic the effective permeability tensor than a change of the intrinsic values of the parameters of the magnetic inclusions. For example, the static effective permeability of a demagnetized medium is given by 
\begin{equation}
\langle\mu_e'(0)\rangle=\frac{1+a\mu_{dw}+(\mu-\mu_{dw})(1-a)+\sqrt{((\mu-\mu_{dw})(1-a))^2+4(1+a)^2\mu\mu_{dw}}}{3(1+a)}
\end{equation}
where $\mu=1+1/\eta$, $\mu_{dw}=1+1/\eta_{dw}$ are the corresponding static permeabilities. It contains an equal contribution of the domain-walls and uniform spin rotation in contradiction to the classical interpretation given by Snoek \cite{Smit:1959df} which links the initial permeability to domain-wall susceptibility only. This formula also contains the fraction of the domain-walls $a$ as a continuous parameter. For large value of $a$, the initial permeability is well described by the permeability of the domain-wall in agreement with the classical interpretation. On the other side, when $a\rightarrow 0$ this is the static part of the domain permeability that drives the initial permeability. The classical interpretation has already been questioned \cite{Bouchaud:1990ty} and the assumption of a perfect addition of the two types of magnetic processes as described in reference \cite{Nakamura:1994zp}, reveals itself puzzling with a varying remanent field.

\begin{figure}[htb]
\resizebox{0.9\columnwidth}{!}{\includegraphics{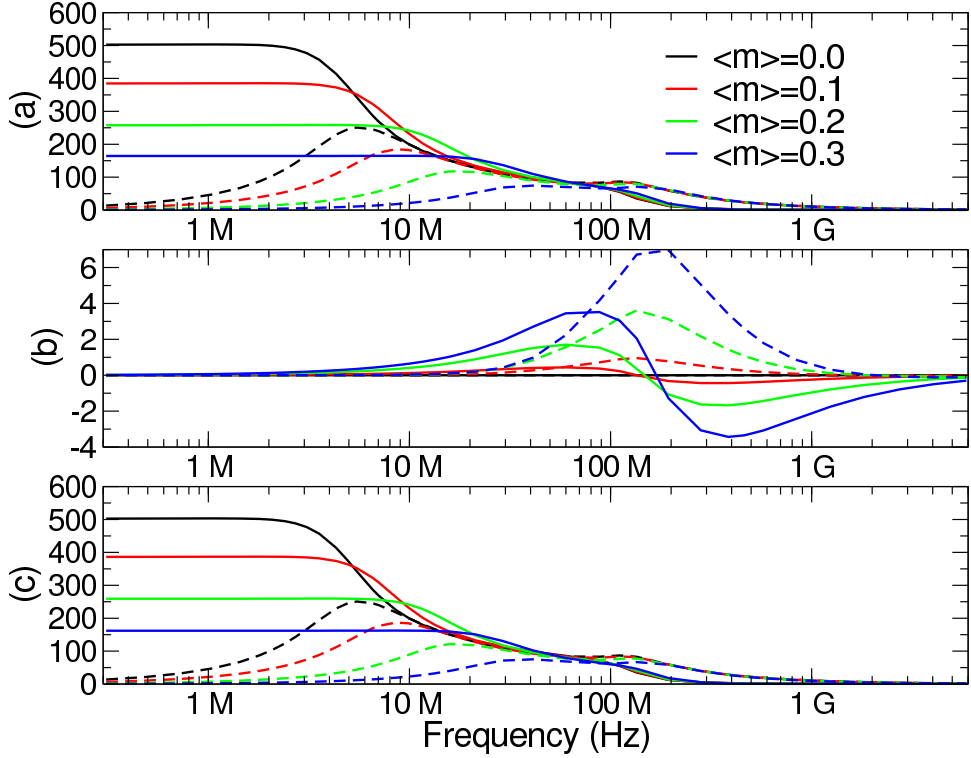}}
\caption{(Color online) Spectral real (straight line) and imaginary (dotted line) part of the effective permeability $\langle{\mu_e}\rangle$ (sub-figure a), effective off-diagonal susceptibility $\langle\kappa_e\rangle$ (sub-figure b) and effective zz permeability $\langle{\mu_e}_z\rangle$ (sub-figure c) as a function of several macroscopic remanent magnetization values $\langle m\rangle$. \label{mu_m_eff}}
\end{figure}

The behavior of the effective diagonal and off-diagonal permeability with an increasing value of the remanent magnetization $\langle m\rangle$ is now investigated. For the same parameters in the table \ref{param_mod}, the spectra are depicted on figure \ref{mu_m_eff}. 
The magnetic behavior of these spectra is strongly similar to those measured from reference \cite{Tsutaoka:1997la}, once given the magnetization law that connects $\langle m\rangle$ to the external uniform magnetic field $H$. Because the domain-wall permeability tensor does not carry any off-diagonal expression, the resulting effective off-diagonal permeability is small at very low frequency and simply dominated by the susceptibly coming from the domains. This is in agreement with first direct observation of domain rotation in ferrites \cite{Brown:1955eu} and by accurate measurements given by Green {\it et al.}\cite{Green:1974qf}.

\section{Conclusion}
Effective permeability tensor for unsaturated polycrystalline ferrites are derived through an EMA and combines domain-wall motion and rotation of domains in a single consistent framework. The dispersion of the local magnetization axis that encodes the polycrystalline character is taken into account by averaging the free energy to restore the magnetic anhysteretic behavior. The initial permeability is given as a mixture formula of magnetic inclusions and fraction of domain-wall which gives a picture of the low-frequency permeability spectra as a magnetic scattering in geometrically arrangement of domains and domain-walls instead of varying material properties. It is envisioned that this theory can be extended to treat multiple scattering of electromagnetic fields in such cylindrical geometry to include multiply peaked spectra observed in several uniform, polycrystalline materials.
\begin{acknowledgments}
JT acknowledges financial support through a joint doctoral fellowship ``CEA-Région Centre''.   
\end{acknowledgments}

\appendix

\section{EMA formalism\label{appendix1}}
The double overbars indicate a tensor notation of rank 2, such as the $i$-th an $j$-th component of a tensor $\overline{\overline{A}}$ is $[\overline{\overline{A}}]_{ij}\equiv A_{ij}$. 
A medium of volume $V$, bounded by a surface $S$ is characterized by a permeability tensor $\overline{\overline{\mu}}$ as a function of frequency and whose values are random in position. It is random in a sense that the configuration average of $\overline{\overline{\mu}}(\vec{r})$, denoted $\langle\overline{\overline{\mu}}\rangle$, is independent of $\vec{r}$. It is also implicitly supposed that this configuration average is equivalent to a volume average, which is the zero wave-vector of the spatial Fourier transform of $\overline{\overline{\mu}}(\vec{r})$, such as $\langle\overline{\overline{\mu}}\rangle\equiv V^{-1}\int_V\overline{\overline{\mu}}(\vec{r})d\vec{r}$. The measurable permeability tensor $\overline{\overline{\mu}}_e$ is the constant of proportionality between  applied magnetic field $\vec{H}_0$ such as $\langle\vec{H}\rangle\equiv\vec{H}_0$ and the average induction $\langle\vec{B}\rangle\equiv\overline{\overline{\mu}}_e\langle\vec{H}\rangle$. Consider that the discussion is restricted to the quasi-static limit when $d/\lambda\ll 1$; $d$ is a typical dimension of the scatterer and $\lambda$ is the wavelength of the incident oscillating magnetic field for that only the lowest order of scattering wave mode is evaluated for high-resistive ferrite \cite{Marysko:1978rw}. In this limit, the magnetic field equations can be approximated by magneto-static equations only
\begin{equation}
\begin{array}{rcl}
\vec{\nabla}\cdot(\mq{\mu}(\vec{r})\vec{H}(\vec{r}))&=&0\\
\vec{\nabla}\times\vec{H}(\vec{r})&=&\vec{0}
\end{array}
\end{equation}    
which are combined up to a given gauge, to find a magnetic potential $\phi(\vec{r})$ inside the medium satisfying 
\begin{equation}
\vec{\nabla}\cdot(\mq{\mu}(\vec{r})\vec{\nabla}\phi(\vec{r}))=0.
\end{equation} 
Now a magnetic tensor $\mm{0}$ is considered to depend on the magnetic field in the average medium. In the quasi-static limit, the boundary conditions of a uniform applied field are imposed and $\mm{0}$ does not exhibit any spatial variations. The local permeability tensor, thus decomposed as $\mm{}(\vec{r})=\mm{0}+\delta\mm{}(\vec{r})$, leads to the following boundary-value problem
\begin{equation}
\begin{array}{rcl}
\vec{\nabla}\cdot(\mm{0}\vec{\nabla}\phi(\vec{r}))&=&-\vec{\nabla}\cdot(\delta\mm{}(\vec{r})\vec{\nabla}\phi(\vec{r}))\textrm{\hspace{5mm}in V},\\
\phi(\vec{r})&=&-\vec{H}_0\cdot\vec{r}\textrm{\hspace{5mm}on S}.
\end{array}
\end{equation}
With the introduction of the two-points Green's function $g(\vec{r},\vec{r'})$ defined by
\begin{equation}
\begin{array}{rcl}
\vec{\nabla}\cdot(\mm{0}\vec{\nabla}g(\vec{r},\vec{r'}))&=&-\delta(\vec{r}-\vec{r'})\textrm{\hspace{5mm}in V},\\
g(\vec{r},\vec{r'})&=&0\textrm{\hspace{5mm}$\vec{r'}$ on S},\label{Green}
\end{array}
\end{equation}
the potential $\phi(\vec{r})$ admits a formal solution as 
\begin{equation}
\phi(\vec{r})=-\vec{H}_0\cdot\vec{r}-\int_Vg(\vec{r},\vec{r'})\vec{\nabla}'\cdot(\delta\mm{}(\vec{r'})\vec{\nabla}'\phi(\vec{r'}))d\vec{r'}
\end{equation}
and the magnetic field $\vec{H}(\vec{r})=-\vec{\nabla}\phi(\vec{r})$, deriving of it, is given by
\begin{equation}
\begin{array}{rcl}
\vec{H}(\vec{r})&=&\vec{H}_0-\int_V(\delta\mm{}(\vec{r'})\vec{H}(\vec{r'}))\vec{\nabla}g(\vec{r},\vec{r'})d\vec{r'},\\
&=&\vec{H}_0+\int_V\mq{G}(\vec{r},\vec{r'})(\delta\mm{}(\vec{r'})\vec{H}(\vec{r'}))d\vec{r'},
\end{array}
\end{equation}
with
\begin{equation}
G_{\alpha\beta}(\vec{r},\vec{r'})=\frac{\partial^2g(\vec{r},\vec{r'})}{\partial r'_\alpha\partial r_\beta}.
\end{equation}
By combining these equations, the problem thus reduces to the task of computing a susceptibility tensor $\langle\mq{\chi}\rangle$ such as
$\delta\mm{}(\vec{r})\vec{H}(\vec{r})=\mq{\chi}(\vec{r})\vec{H}_0$ and 
\begin{equation}
\mq{\chi}(\vec{r})=\delta\mm{}(\vec{r})\left(\mq{1}+\int_V\mq{G}(\vec{r},\vec{r'})\mq{\chi}(\vec{r'})d\vec{r'}\right),\label{EMTCHI}
\end{equation}
which finally gives
\begin{equation}
\mm{e}=\mm{0}+\langle\mq{\chi}(\vec{r})\rangle.
\end{equation}
If $\vec{r}$ lies in a medium labeled $i$, of volume $v_i$, equation (\ref{EMTCHI}) decomposes itself as
\begin{eqnarray}
\mq{\chi}(\vec{r})&=&\delta\mm{i}\left(\mq{1}+\int_{v_i}\mq{G}(\vec{r},\vec{r'})\mq{\chi}(\vec{r'})d\vec{r'}\nonumber+\int_{V-v_i}\mq{G}(\vec{r},\vec{r'})\mq{\chi}(\vec{r'})d\vec{r'}\right),
\end{eqnarray}
with $\delta\mm{i}\equiv\mm{i}-\mm{0}$, and $\mm{i}$ is the permeability tensor of the medium $i$, assumed spatially uniform and known. The last integral is approximated by 
\begin{equation}
\int_{V-v_i}\mq{G}(\vec{r},\vec{r'})\mq{\chi}(\vec{r'})d\vec{r'}\approx\int_{V-v_i}\mq{G}(\vec{r},\vec{r'})\langle\mq{\chi}(\vec{r'})\rangle d\vec{r'}
\end{equation}
once neglecting higher order correlation terms. This last expression closes the integral equation for $\mq{\chi}(\vec{r})$. To see this, this last integral is substituted, integrated by parts and by imposing the boundary condition one obtains
\begin{equation}
\mq{\chi}_i=\left(\mq{1}-\delta\mm{i}\mq{\Gamma}_i\right)^{-1}\delta\mm{i}\left(\mq{1}-\mq{\Gamma}_i\langle\mq{\chi}\rangle\right)
\end{equation}
where $\mq{\chi}_i$ stands for $\mq{\chi}(\vec{r})$ for $\vec{r}\in v_i$. Here $\mq{\Gamma}_i$ is a surface integral, also called depolarization matrix and contains $g(\vec{r},\vec{r'})$ which goes over to the free-space Green's function, satisfying the differential equation (\ref{Green}) and the boundary condition $g(\vec{r},\vec{r'})\rightarrow 0$ as $\|\vec{r}-\vec{r'}\|\rightarrow\infty$. By reason of the translational invariance, it becomes a function of $\vec{r}-\vec{r'}$ to finally reduces to a constant in space variables. For example, the field inside ellipsoids is uniform and $\mq{\chi}(\vec{r})$ becomes independent of the position. In such case, each cartesian component of this matrix is given by 
\begin{equation}
\label{gammatensor}
\Gamma^{\alpha\beta}_i=-\oint_{S'}\frac{\partial g(\vec{r}-\vec{r'})}{\partial r_\alpha} n'_\beta d^2\vec{r'},
\end{equation}
where $n'_\beta$ is the component of the unit normal outward from the surface $S'$. By taking the average of $\mq{\chi}_i$, one has
\begin{equation}
\mm{e}=\mm{0}+\langle(\mq{1}-\delta\mm{}\mq{\Gamma})^{-1}\rangle^{-1}\langle(\mq{1}-\delta\mm{}\mq{\Gamma})^{-1}\delta\mm{}\rangle\label{CM1}
\end{equation}
where
\begin{eqnarray}
\langle(\mq{1}-\delta\mm{}\mq{\Gamma})^{-1}\rangle&\equiv&\displaystyle{\lim_{V\rightarrow\infty}\sum_iv_i(\mq{1}-\delta\mm{i}\mq{\Gamma}_i)^{-1}}\label{EMT1a}\\
\langle(\mq{1}-\delta\mm{}\mq{\Gamma})^{-1}\delta\mm{}\rangle&\equiv&\displaystyle{\lim_{V\rightarrow\infty}\sum_iv_i(\mq{1}-\delta\mm{i}\mq{\Gamma}_i)^{-1}\delta\mm{i}}\label{EMT1b}
\end{eqnarray}
Now, a two-phase medium of magnetic objects labeled by $i=1,2$, each fully characterized by their respective position and field independent permeability tensors $\overline{\overline{\mu}}_i$ is considered. If $\mm{1}$, $\mm{2}$, $\mm{e}$ and $\mq{\Gamma}_{1,2}=\mq{\Gamma}$ are all hermitian matrices, and if $\mm{1}=\mm{0}$, one shows that equation (\ref{CM1}) is equivalent to the Clausius-Mossotti-Maxwell-Garnett expression \cite{Maxwell-Garnett:1904cs}
\begin{equation}
(\mq{1}-(\mm{e}-\mm{1})\mq{\Gamma})^{-1}(\mm{e}-\mm{1})=f(\mq{1}-(\mm{2}-\mm{1})\mq{\Gamma})^{-1}(\mm{2}-\mm{1})\label{CM2}
\end{equation}
with $f$ is the volume fraction of object $2$.

\bibliographystyle{apsrev4-1}
\bibliography{unsatfer}

\end{document}